# A Quantum-Chemical Simulation of the Cadmium Chalcogenide Clusters Terminated with Hydrogen and Simple Functional Groups

V. S. Gurin

Physico-Chemical Research Institute, Belarusian State University, Leningradskaya str., 14, Minsk, 220050, Belarus

**Abstract**

*Ab initio* calculations of cadmium chalcogenide nanoclusters with terminating groups (H, O, N, and C atoms bound to the surface sulfur) are considered as a simulation for the species produced in typical synthesis of bio-conjugates with luminescent quantum dots. The approaches based on the Hartree-Fock (HF) method and density functional theory (DFT) are used assuming geometry optimization keeping the tetrahedral symmetry. The geometry and electronic structure of CdX (X=S, Se, Te) clusters with size up to $Cd_{17}$ (HF) and $Cd_4$ (DFT) bound with hydrogen and -OH, -$NH_2$, -$CH_3$, and –$CH_2CH_3$ groups are calculated and the effects of the terminating groups upon core clusters are discussed.

# 1. Introduction

A bioinspired application of the luminescence phenomenon of semiconductor nanoclusters is of great interest during last decade [1-4 and Refs therein]. The featured photoluminescence properties of the nanoclusters have been studied very intensively together with many another properties provided by the explicit size-dependency and quantum confinement effects [5-8]. Research on luminescence was combined successfully with a variety of nanoscale design aimed at the creation of novel artificial species by the type of 'quantum dot (QD) – biomolecule'. There are a number of subdirections in research of bio-conjugated QD, however, one of most popular applications to date is associated with their striking optical properties of QD and consists, basically, in designing of luminescent labels of biomolecules, in particular, DNA. The 'labeling' assumes the usage of features of both QD and biomolecules without their destroy and with minimum effect upon them in order to keep both luminescence properties of QD and the function of molecules. Experimental findings in this field are quite successful to date due to very intensive research efforts [1-8]. On the other hand, theory of QD-biomolecule interaction is rather restricted because of great complicacy of a complete task for realistic species. Surface chemistry of semiconductor nanoclusters, in contrast to bulk counterparts, is very flexible and poorly investigated in detail to date since a usage of direct surface investigating techniques is problematic for tiny boundary areas of nanoparticles. As well the bulk surface, nanoclusters possess the strong surface reactivity due to dangling bonds at crystal (nanocrystal) boundaries, appearance of which is principal for any consideration of solids under diminishing of dimensions. However, a contribution of surface groups is known to be much larger at nanometer scales, and the surface features of semiconductor nanoclusters are of great importance for their properties. The linking with some molecules is one such property. A general principle of QD-molecule formation consists in a surface reaction of molecules with terminating groups of nanoclusters, and has been realized for many systems using, dominantly, II-VI QDs as strongly light-emitting labels. Principal constituents of the surface bonding can be simulated at different theory level to shed a light onto the nature of the surface functionality of QD with respect to organic molecules. This work is devoted to construction and calculation of possible models of surface bonding with semiconductor nanoclusters.

Cadmium chalcogenides occupy a challenging place among nanoscale semiconductors of the different chemical nature due to popularity in both experimental studies and applications [5-8]. A number of studies are devoted to cadmium chalcogenide clusters of small size ranging from 2 to several tens of atoms [9-29]. Experimentally, a variety of structures with $Cd_nX_m$ cores have been produced and characterized [9-18], while a theory is extended from the simplest $Cd_2X_2$ models throughout the clusters of more than 100 metal atoms [19-31]. An ability to calculate small or large clusters is motivated by approximations used in the corresponding level of theory. However, high-level calculations with adequate account of electronic correlation and excited states are available now only for the small clusters with several metal atoms [25].

In the present work, we consider *ab initio* calculations of cadmium chalcogenide nanoclusters with shells of organic molecules as a working model for the species produced in typical synthesis of DNA-conjugates with QDs. We use both HF and DFT approaches assuming arbitrary geometry optimization and derive the electronic structure and optical properties of the species with size up to several tens of atoms within QDs with organic shells. The theory is restricted to the level allowing to simulate both nanoclusters with tens of heavy atoms and include different terminating group around them under the reasonable computational cost.

## 2. The models under study and calculation details

Geometry of the CdX-cluster cores under study was built as the fragments of bulk cadmium chalcogenide lattice of the sphalerite type, and the tetrahedral coordination of both Cd and S atoms was retained. The tetrahedral symmetry was kept explicitly for generation of models and for reduction of computational resources. Such approach assumes isotropic environment of the nanoclusters and free linking the terminating groups. More complicated geometry for the cores can be considered in future in the analogous manner. This choice of these $Cd_nX_m$ cluster models has been motivated by the knowledge on structures of nanocluster those are really synthesized in liquid medium with protecting groups [32,33], however, there are another versions of clusters with n and m providing the rules of tetrahedral coordination of Cd and S [24], and in some cases with another environment (gaseous phase, vacuum) the situation with structure and elemental composition can be also different[14].

Fig. 1 displays the calculation data for $Cd_{13}X_{16}$ clusters which can be considered as built through the sequence of $CdX_4$–$Cd_{13}X_4$ - $Cd_{13}X_{16}$ starting from one central Cd atom by coordinating one shell of X atoms, then joining more Cd atoms followed by the next shell of X atoms. The further joining of Cd atoms followed by the next X shell to the externals of $Cd_{13}X_{16}$ results in $Cd_{17}X_{28}$ clusters (Fig. 1). External chalcogen atoms in these clusters were terminated by hydrogen atoms in order to simulate the simplest protecting groups of real clusters. The models in DFT calculations were based on the constuction with $Cd_4S_6$ cores (Fig. 2) and termination by hydrogen atoms and a series of other groups was considered.

The calculation methods used within the framework of this work include both restricted HF (RHF) and DFT schemes. Basis sets for Cd and X atoms were with effective core potentials for internal core orbitals (28 electron core for Cd and 10, 28, 46-electron core for S, Se, Te, respectively) designed by Hay and Wadt (LANL2DZ in the Gaussian notation) [35] that shows good choice in many cases for the elements of medium part of the Periodic system. The lighter atoms (H, C, N, O) were calculated with all-electron 6-31G basis sets. Singlet states (closed shell) were taken as the lowest energy ones for all models under study in this work and the geometry optimization was attained for them. GAMESS(US) [36] and NWChem 4.5 [37] packages were utilized for calculations.

## 3. Results and discussion of the cluster properties

### 3.1. RHF calculations of clusters with $Cd_{13}S_{16}$- and $Cd_{17}S_{28}$-cores

The main calculation data are presented in Fig. 1 for the optimized geometries with effective charges at atoms those are unique within the symmetry group. Table 1 collects the values of energy of HOMO-LUMO (highest occupied - lowest unoccupied molecular orbitals) transitions. The HOMO-LUMO gap is the first estimation for the interband transition energies.

According to the calculation data for $Cd_{13}S_{16}H_{12}^{6+}$ cluster, the different values of Cd-S distances occur if to compare internal atoms and the atoms near its surface. The minimum distance appears for the second layer, and this value is less than the Cd-S distance in the bulk sphalerite CdS (2.52 Å). The other Cd-S bonds in this cluster are longer than the bulk value in consistence with experimental observation [38]. These data show that an assumption of the bulk cluster geometries without correction of the finite size effect is not very good approximation, and there exists an explicit surface effect. The bond with hydrogen atoms, S-H, is longer than this bond length in $H_2S$ molecule (1.336 Å), but this difference is not big. Thus, the hydrogen termination can simulate quite well other bonding with larger moieties.

The values of effective charges at atoms in the calculated models demonstrate the featured distribution of electronic density within clusters. Evidently, all Cd and S atoms possess same charges in the bulk CdS, however, very inhomogeneous distribution of charges occurs in the clusters: internal sulfur atoms possess the charge –0.91 and external ones do –0.46. The similar difference exists also for the clusters with selenium and tellurium. As the whole, the internal bonds are more ionic than the surface bonds. Hydrogen atoms keep also the rather high charges, that corresponds to the rather strong effect of hydrogen termination upon the state of clusters. Other groups in similar manner can interact with surface atoms and stabilize them due to redistribution of extra electronic density.

The clusters built on the basis of CdSe and CdTe with similar composition, $Cd_{13}Se_{16}H_{12}^{6+}$ and $Cd_{13}Te_{16}H_{12}^{6+}$, discover remarkable variance in the calculated characteristics as compared with CdS-ones. In spite of very similar geometries, the clusters in CdS-CdSe-CdTe sequence are expected to be different both in reactivity and electronic properties. Selenide and telluride clusters show the longer Cd-X bonds as expected due to the larger atomic radii of Se and Te, and Cd-Te bonds are longer than Cd-Se ones. However, the minimum bond length of Cd-Se and Cd-Te are observed for the near-surface atoms rather than the internal second layer in $Cd_{13}S_{16}H_{12}^{6+}$. Distribution of charges in selenide and telluride clusters is more uniform than in sulfide, and ionity of bonds is lower. In the case of $Cd_{13}Te_{16}H_{12}^{6+}$ some surface charges appear to be even inverted, so, tellurium atoms become slightly positive. These variations do not contradict to general expectations on properties of atoms in S-Se-Te series. It is of interest to note that hydrogen atoms possess very low effective charges in the case of $Cd_{13}Te_{16}H_{12}^{6+}$, and the corresponding bond length, X-H, is remarkably longer than the bond length in $Cd_{13}S_{16}H_{12}^{6+}$. The latter circumstance can mean that the role of surface termination diminishes noticeably going from sulfide to selenide and telluride clusters since hydrogen atoms weakly bond with the core. Thus, an existence of stable 'naked' $Cd_{13}X_{16}$ clusters in the case of telluride is more probable, and different surface groups can have the less effect upon it than in the case of $Cd_{13}Se_{16}$ and $Cd_{13}S_{16}$. On the other hand, a linking with any terminating group will results in the minimum changes of telluride clusters and the maximum ones will be upon the sulfides.

The larger clusters under calculation here, $Cd_{17}X_{28}H_{24}^{2+}$, are of interest from the point of view of experimental verification, as far as there exist the species of the dimeric composition $[Cd_{17}S_{30}R_{26}]_2$, where terminating group R = -$CH_2CH_2OH$. The theorerical structure of the corresponding core $Cd_{17}S_{28}$ presents the tetrahedron unit of the crystal lattice established in [33]. Here we simulate the termination of $Cd_{17}X_{28}$ cores by hydrogen atoms (Fig. 1). In the case of $Cd_{17}S_{28}H_{24}^{2+}$ the Cd-S distances are shorter for internal part, and essentially longer for the external one, similar to the situation with the above $Cd_{13}S_{16}H_{12}^{6+}$. However, for $Cd_{17}Te_{28}H_{24}^{2+}$ this comparison of distances is reversed (internal ones are longer) and for $Cd_{17}Se_{28}H_{24}^{2+}$ all Cd-X distances are almost of the same value (Fig. 1). Thus, the geometrical parameters of the selenide clusters appear to be closest to the bulk state, and any protecting groups are to be of minimum effect upon the $Cd_{17}$-selenide than on the corresponding sulfide and telluride. The effective charges for surface chalcogen atoms are almost the same for $Cd_{17}Se_{28}H_{24}^{2+}$ and $Cd_{17}Te_{28}H_{24}^{2+}$, but the sulfide counterpart appears to be more polarized.

The electronic transitions simulated here through the HOMO-LUMO gap for these clusters with $Cd_{13}$- and $Cd_{17}$-cores evidence also on the noticeable effects due to the type of X atoms. Also, a comparison of the $Cd_{13}$- and $Cd_{17}$ clusters reveals the reasonable variation in of this gap which is decreased on more than 2 eV for $Cd_{17}$- as compared with $Cd_{13}$- species. The value for the $Cd_{17}$- approaches to the first absorption maximum of the nanoclusters produced experimentally [33]. However, it remains to be overestimated if to assign to ~1 nm clusters that can be due to deficiency of this model with H- termination to reproduce exactly the termination of more complex groups.

### 3.2. DFT calculations of clusters with $Cd_4S_6$-cores

The optimized geometry of these models is presented in Fig. 2. They were built on the basis of $Cd_4S_6$ core terminating the sulfur atoms by hydrogen and the other groups resulting in formation of S-H, S-C, S-O, and S-N bonds. The calculation results indicate clearly that both geometry and electron density distribution are changed dramatically in dependence on the type of termination groups. That may be understandable due to small size of the clusters with four Cd atoms and properties of the cores are far from those of the bulk CdX counterparts.

The $Cd_4S_6$-$H_6^{2+}$ model, i.e. the core with the simplest version of termination, demonstrates the equilibrium Cd-S bond length very close to the bulk CdS value, however in the case of another terminating groups (below) this value is seen to be strongly changed. The remarkable variations occur also for the bond lengths to the heretoatoms: the maximum one is obtained in the case of oxygen, S-O, that is associated, evidently, with different atomic radii of O, N, and C atoms, respectively.

The termination with hydroxide group, in the case of $Cd_4S_6$-$(OH)_6^{2+}$, changes the features of clusters dramatically, and the value of Cd-S bond length increases noticeably together with strong distortion of the cluster core. This also accompanied by decrease of Cd-S bond ionity. The geometry for termination with amine group, $Cd_4S_6$-$(NH_2)_6^{2+}$, is less distorted than for hydroxide, however, the perturbation upon the core structure is also rather essential with decreased Cd-S bond ionity. The strong effects due to both hydroxide- and amine-termination are understandable based on the known powerful polarization ability of these groups. Meanwhile, there is rather unexpected effect of the methyl group upon $Cd_4S_6$ core observed by the next model calculated

(Fig. 2). That can be reasoned due to the large size of this group resulting in steric effects of three hydrogen atoms at each carbon atom. Thus, a termination by this 1-C groups cannot be considered as a stabilizing factor for $Cd_4S_6$ clusters. However, the more hopeful situation was obtained for the termination by the 2-C group, in the case of $Cd_4S_6\text{-}(CH_2\text{-}CH_3)_6^{2+}$ model. The core geometry distortion is seen to be much less than the distortion with lighter terminating group (besides the H-termination case). This model with the 2-C terminating group can be assigned as the maximum stabilizing one. The terminating groups in the above calculations can be considered as elementary models of the bonds with more complicated organic species.

## 4. Conclusions

1. CdX nanoclusters have been simulated with the terminating groups by means of *ab initio* quantum chemistry in order to consider simplest versions of linking with organic molecules.
2. RHF calculations of the clusters with 13 and 17 Cd atoms reveal the variation of the cluster parameters (interatomic distances, effective charges, HOMO-LUMO transition energy) in dependence on number of atoms, surface termination, and the chemical nature of chalcogen atoms. The atoms in the clusters are not equivalent from the depth to the surface.
3. DFT calculations of the clusters with $Cd_4S_6$-core demonstrate the pronounced effect due to different terminating groups. –H, -OH, -$NH_2$, -$CH_3$, and -$CH_2CH_3$ were tested, and the groups with one non-H reveal the strong effect upon geometry of the core, while the double-carbon group indicates minimum changes both in the electronic structure and geometry of $Cd_4S_6$.

Table 1.

Calculated energies of dipole allowed HOMO-LUMO transitions of the $Cd_{13}$- and $Cd_{17}$- clusters

| Cluster | Energy, eV | Symmetry of HOMO and LUMO |
|---|---|---|
| $Cd_{13}S_{16}H_{12}^{6+}$ | 8.90 | $t_2 \rightarrow a_1$ |
| $Cd_{13}Se_{16}H_{12}^{6+}$ | 8.49 | $t_1 \rightarrow t_2$ |
| $Cd_{13}Te_{16}H_{12}^{6+}$ | 8.17 | $t_1 \rightarrow t_2$ |
|  |  |  |
| $Cd_{17}S_{28}H_{24}^{2+}$ | 6.69 | $t_2 \rightarrow t_1$ |
| $Cd_{17}Se_{28}H_{24}^{2+}$ | 5.88 | $t_2 \rightarrow a_1$ |
| $Cd_{17}Te_{28}H_{24}^{2+}$ | 5.89 | $a_2 \rightarrow a_1$ |

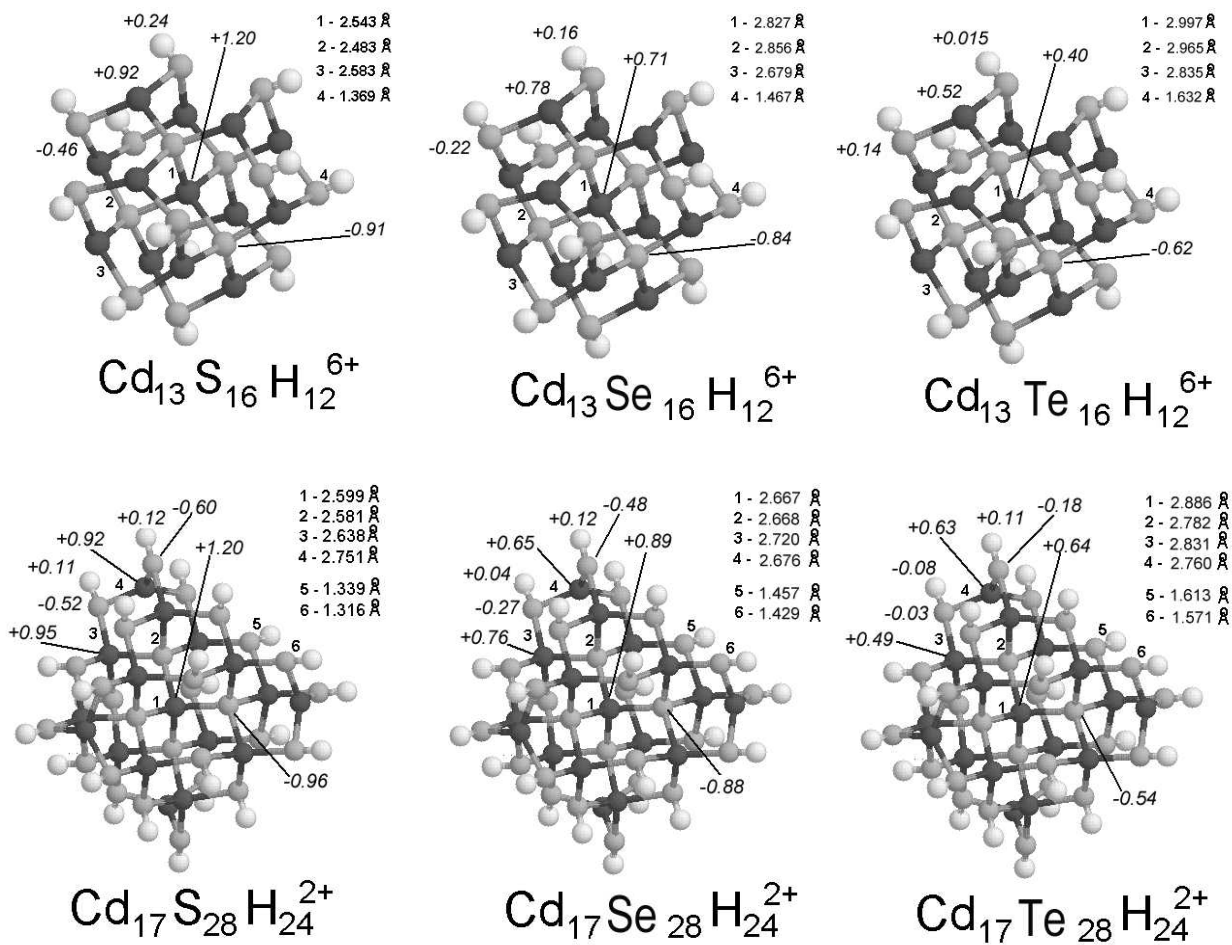

**Fig. 1.** Geometry of (a) $Cd_{13}X_{16}$- and (b) $Cd_{17}X_{28}$- clusters calculated at the HF level.

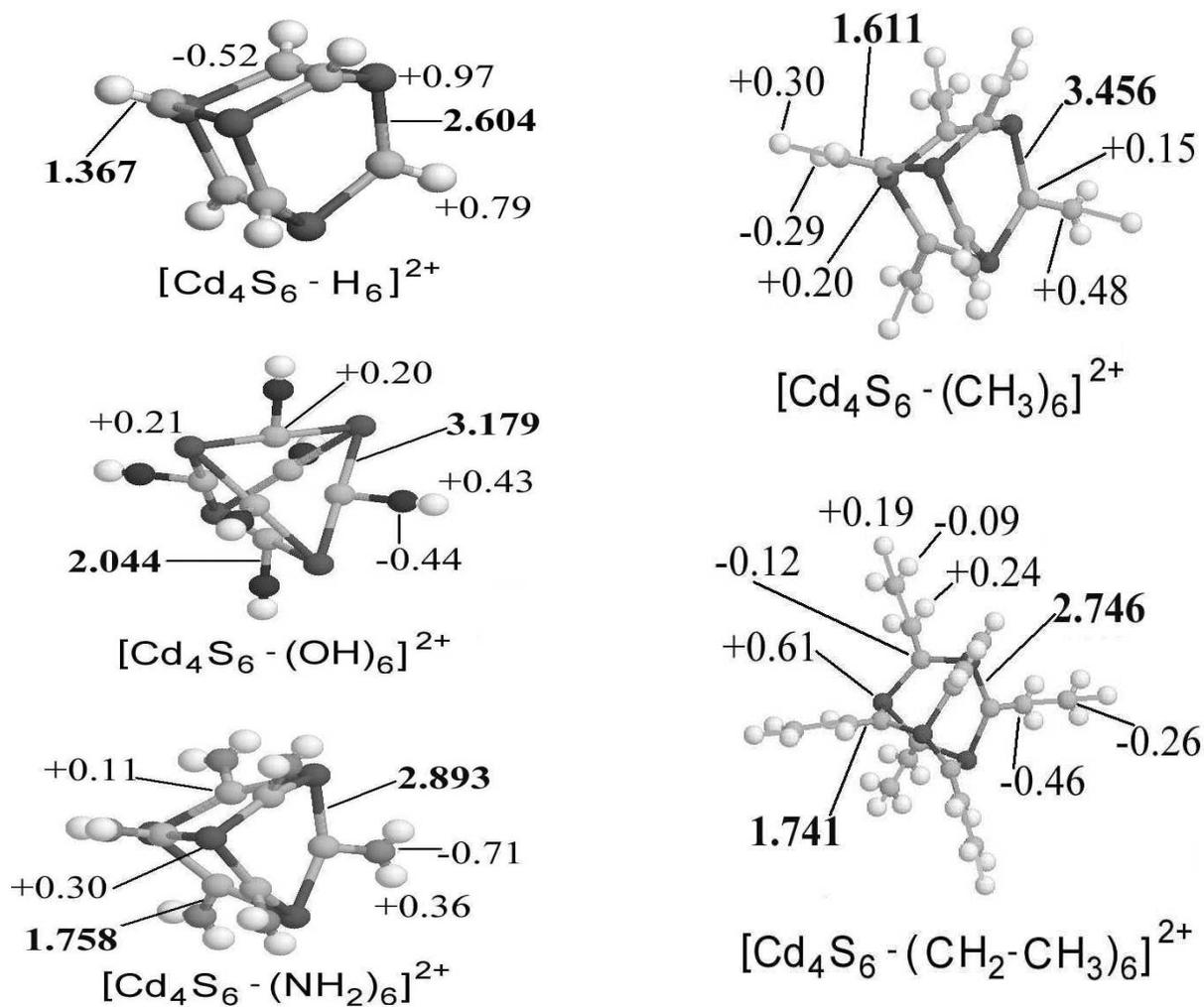

**Fig. 2.** Geometry of a series of $Cd_4S_6$- clusters calculated at the DFT level with effective charges at the atoms (regular font) and interatomic distances Cd-S, S-H, S-C, S-O, and S-N (Å, bold face).

# References

[1] E. Katz, A.N. Shipway, and I. Willner. In: Nanoparticles. G. Schmid. (Ed.), Wiley-VCH Verlag GmbH& Co. KGaA, WeinHeim. 2004. pp. 368-421.

[2] L. Pasquato, P. Pengo, and P. Scrimin, In: Nanoparticles. Building Blocks for Nanotechnology. V. Rotello (Ed.). Kluwer Acad. Plenum Publ., New York, etc. 2003. pp. 251-281.

[3] V. Biju, T. Itoh, A. Anas, A. Sujith, and M. Ishikawa. Anal. Bioanal. Chem. 391 (2008) 2469-2495.

[4] Nanofabrication Towards Biomedical Application. Ch.S.S.R. Kumar, J. Hormes, C. Leuschner (Eds). Wiley-VCH Verlag GmbH& Co. KGaA, WeinHeim. 2005.

[5] S.V. Gaponenko, Optical properties of semiconductor nanocrystals, Cambridge University Press, Cambridge, 1998.

[6] Semiconductor Nanocrystal Quantum Dots. A. Rogach (Ed.). SpringerWienNewYork. 2008.

[7] U. Woggon, Optical properties of semiconductor quantum dots, Springer-Verlag, Berlin, 1998.

[8] C. Delerue and M. Lannoo. Nanostructures. Theory and Modeling. Springer-Verlag, Berlin, Heidelberg, 2004.

[9] B. Krebs and G. Henkel, Angew. Chem. Int. Ed. Engl. 30 (1991) 769-788.

[10] I.G. Dance, Progr. Inorg. Chem. 41 (1994) 637-803.

[11] J.-O. Joswig, G. Seifert, T.A. Niehaus, and M. Springborg, J. Phys. Chem. B 107 (2003) 2897-2892.

[12] Th. Lover, G.A. Bowmaker, J.M. Seakins, and R.P. Cooney, Chem. Mater. 9 (1997) 967-975.

[13] G.S.H. Lee, K.J. Fisher, A.M. Vassallo, J.V. Hanna, and I.G. Dance, Inorg. Chem. 32 (1993) 66-72.

[14] A. Kasuya, R. Sivamohan, Y.A. Barnakov et al, Nature Mater. 3 (2004) 99-102.

[15] H.-J. Liu, J.T. Hupp, and M.A. Ratner, J. Phys. Chem. 100 (1996) 12204-12213.

[16] A. Jentys, R.W. Grimes, J.D. Gale, C.R.A. Catlow, J. Phys. Chem. 97 (1993) 13535-13538.

[17] H. Hosokawa, H. Fujiwara, K. Murakoshi, Y. Wada, Sh. Yanagida, and M. Satoh, J. Phys. Chem. 100 (1996) 6649-6656.

[18] D. D. Lovingood, R.E. Oyler, and G.F. Strouse, J. Am. Chem. Soc. 130 (2008) 17004-17011.

[19] M.C. Troparevsky, L. Kronik, and J.R. Chelikowsky, J. Chem. Phys. 119 (2003) 2284-2287.

[20] K. Eichkorn and R. Ahlrichs, Chem. Phys. Lett. 288 (1998) 235-242.

[21] P. Deglmann, R. Ahlrichs, and K. Tsereteli, J. Chem. Phys. 116 (2002) 1585-1597.

[22] J. M. Matxain, J. M. Mercero, J.E. Fowler, and J.M. Ugalde, J. Phys. Chem. A 108 (2004) 10502-10508.

[23] V.N. Soloviev, A. Eichhoefer, D. Fenske, and U. Banin, J. Am. Chem. Soc. 123 (2001) 2354-2364.

[24] P. Sarkar and M. Springborg, Phys. Rev. B 68 (2003) 235409/1-235409/7

[25] P. Yang, S. Tretiak, A.E.Masunov, and S. Ivanov, J. Chem. Phys. 129 (2008) 074709.

[26] Sh. Xu, Ch. Wang, and Y. Cui, J. Mol. Model. (2009) DOI: 10.1007/s00894-009-0564-4

[27] O.V. Prezhdo, Chem. Phys. Lett. 460 (2008) 1-9.

[28] J. Frenzel, J.-O. Joswig, and G. Seifert, J. Phys. Chem. C 111 (2007) 10761-10770.

[29] G. Maroulis and C. Pouchan, Chem. Phys. Lett. 464 (2008) 16-20.

[30] V.S. Gurin, Intern. J. Quantum Chem. 71 (1999) 337-341.

[31] V.S. Gurin, Surf. Rev. Lett. 7 (2000) 161-169.

[32] Y. Nosaka, H. Shigeno, and T. Ikeuchi, J. Phys. Chem. 99 (1995) 8317-8322.

[33] T. Vossmeyer, G. Reck, L. Katsikas, E.T.K Haupt,.B. Schulz, and H. Weller, Science 267 (1995) 1476-1479.

[34] V.S. Gurin, Colloid and Surfaces:A: 202 (2002) 215-222.

[35] P.J. Hay and W.R. Wadt, J. Chem. Phys. 82 (1985) 270-283.

[36] M.W. Schmidt, K.K. Baldridge, J.A. Boatz, et al., J. Comput. Chem. 14 (1993) 1347-1363.

[37]. R. Harrison, J.A. Nichols, T.P. Straatsma, M. Dupuis et al. NWChem, Computational Chemistry Package for Parallel Computers, Version 4.5. 2004, Pacific Northwest National Laboratory, Richland, Washington 99352-0999, USA.

[38] J. Rockenberger, L. Troger, A. Kornowski, T. Vossmeyer, A. Eychmueller, J. Feldhaus, and H. Weller, J. Phys. Chem. B101 (1997) 2691-2701.